\documentclass[twocolumn,aps,prb]{revtex4}
\usepackage{graphicx,graphics,color,epsfig}
\usepackage{amsmath}
\usepackage{amssymb}
\usepackage[small]{caption}

\begin{document}
\title{
Finite spin-glass transition of the $\pm J$ XY model in three dimensions
}
\author{Takeo Yamamoto, Takeshi Sugashima and Tota Nakamura}
\affiliation{Department of Applied Physics, Tohoku University,
             Aoba, Sendai, Miyagi 980-8579, Japan}

\date{\today}

\begin{abstract}
A three-dimensional $\pm J$ XY spin-glass model is investigated
by a nonequilibrium relaxation method.
We have introduced a new criterion for the finite-time scaling analysis.
A transition temperature is obtained by a crossing point of obtained data.
The scaling analysis on the relaxation functions of the
spin-glass susceptibility and the chiral-glass susceptibility shows
that both transitions occur simultaneously. 
The result is checked by relaxation functions of the Binder parameters
and the glass correlation lengths of the spin and the chirality.
Every result is consistent if we consider that the transition is
driven by the spin degrees of freedom.
\end{abstract}
\maketitle

\section{Introduction}
\label{sec:intro}

Random and/or frustrated systems exhibit a variety of exotic phenomena.
Common knowledge of uniform systems may not be applied to these systems.
Randomness sometimes brings order to disorder.
Exotic states may appear in the ground state.
New-type phase transitions occur in these systems.
They are referred to as complex systems and have been attracting
wide interest.

Spin glasses \cite{SGReview}
are prototype of complex systems.
Ferromagnetic interaction and antiferromagnetic interaction
are randomly distributed in the materials.
There is frustration in local interaction bonds.
Magnetic spins cannot always minimize the interaction energy.
They freeze at low temperatures while pointing
random directions.
This freezing is called the spin-glass transition.

Theoretical investigations on spin glasses are mainly made into the
Edwards-Anderson model\cite{Edwards}.
Spins are located on a regular lattice but the interaction bonds between
spins are randomly distributed.
Many theoretical methods have been developed.
By progress of computational facilities
numerical simulations are recently performed.
However, simulations in spin glasses suffer from serious slow dynamics.
It takes a long time to realize equilibrium states at low temperatures.
A numerical method overcoming this difficulty can be successfully
applied to other complex systems.
One example is an application of the temperature-exchange Monte
Carlo method\cite{hukuexchange} to protein 
folding problems\cite{okamoto1}.

Not many problems have been made clear 
due to the difficulty of simulations in spin glasses.
One unsolved problem is whether or not a theoretical model with short-range
interaction bonds can explain the real spin-glass transition.
There is a general consensus that the spin-glass transition occurs 
in the Ising model in three dimensions.\cite{Bhatt,Ogielski,KawashimaY}
The estimated critical exponents agree with the corresponding experimental
results\cite{Aruga}.
Spins of many spin-glass materials are of Heisenberg type.
The Heisenberg spin-glass model in three dimensions should exhibit 
the spin-glass transition.
However, numerical investigations \cite{McMillan,Olive,Morris} 
concluded that there is no spin-glass transition in this model.
There has been discrepancy between theory and experiment in this model.

There are three major theories explaining this discrepancy.
One is the anisotropy theory.
There are finite magnetic anisotropies in exchange interactions 
in real materials.
Even though there is no spin-glass order in the isotropic Heisenberg model,
the spin-glass transitions in real materials are explained by finite anisotropy.
Matsubara {\it et al.} showed that the anisotropy reinforces
the spin-glass order.\cite{Matsubara}
The second theory is the chirality mechanism introduced by 
Kawamura.\cite{KawamuraH1,HukushimaH}
It is argued that there occurs the chiral-glass transition without the
spin-glass transition.
Finite anisotropy in real materials mixes the chiral degrees of freedom
and the spin degrees of freedom.
This mixture links the chirality and the spin.
The spin-glass transitions in real materials are driven by this chiral-glass
transition.
This scenario is based on that there is no spin-glass transition in the
isotropic model.
Since the chirality is defined by spin variables, the chiral-glass
transition trivially occurs if the spin-glass transition occurs.
Therefore, it is crucial to check whether both transitions occurs at the
same temperature or not.
The third theory argues a possibility of this simultaneous transition.
Matsubara {\it et al.} \cite{Matsubara3,Matsubara4,Endoh2}
showed several lines of evidence that the spin-glass
transition occurs at a finite temperature.
Nakamura {\it et al.} \cite{Nakamura,Nakamura2}
investigated the model by a nonequilibrium relaxation 
method \cite{Stauffer,Ito,ItoOz1,OzIto2,Shirahata2}
and obtained the consistent results.
Here, we note that
nonequilibrium relaxation of the spin-glass susceptibility\cite{Huse}
and that of the Binder parameter\cite{Blundell} are known to exhibit 
the algebraic divergence at the spin-glass transition temperature
in the Ising model.
Lee and Young \cite{Lee}
also concluded a single chiral-glass and spin-glass transition 
in the model with Gaussian random bond distributions.
These results suggest that the previous investigations which concluded
no spin-glass transition have some technical problems.

The situation is same in the XY spin-glass model in three dimensions.
Spins are restricted to point in the XY plane in this model.
A domain-wall energy analysis \cite{Morris} and a Monte Carlo simulation
analysis \cite{Jain} concluded no spin-glass transition.
The chirality mechanism was originally proposed in this
model.\cite{KawamuraXY1,KawamuraXY2,KawamuraXY3}
However, a possibility of the finite spin-glass transition has been suggested
by recent investigations \cite{Lee,Maucourt,Kosterlitz,Granato,yamamotoissp}.

In this paper we focus on the problem in the XY spin-glass model in three
dimensions: whether or not the spin-glass transition and the chiral-glass
transition occur at the same temperature.
It is also made clear why there has been the discrepancy
even though investigating the same theoretical model with similar methods.
We consider the following three points as origins of the discrepancy.
One is ambiguity in the scaling analysis.
The obtained results are sometimes dependent on a temperature range of data
used in the analysis.
We introduce a new scaling criterion to eliminate this ambiguity.
The second point is strong effects from system sizes and boundary conditions.
We carefully observe how these effects appear in the relaxation data.
It is found that the effects are very strong and strange.
The third point is a use of the spin-glass susceptibility and the Binder
parameter in the equilibrium simulations on finite lattices.
Both quantities are under the strong influence of the size effects.
These points are discussed in this paper.

The present paper is organized as follows.
Section \ref{sec:modelmethod} describes the present model and our method.
A new scaling procedure is explained with an example of the
ferromagnetic Ising model in two dimensions.
Section \ref{sec:results} shows our results.
A finite-time-scaling analysis on the glass susceptibility is performed.
The results are confirmed by relaxation functions of the Binder parameter
and the glass correlation lengths.
Section \ref{sec:discussion} is devoted to summary and discussions.

\section{Model and Method}
\label{sec:modelmethod}

\subsection{Model and physical quantities}
\label{sec:method}
A model we consider in this paper is the $\pm J$ random bond XY model
in a simple cubic lattice.
The Hamiltonian is given by
\begin{equation}
  {\cal H} = - \sum_{\langle ij \rangle} 
     J_{ij} \mbox{\boldmath $S$}_i \cdot 
            \mbox{\boldmath $S$}_j =
 - \sum_{\langle ij \rangle}
     J_{ij} \cos{(\theta_i - \theta_j)} ,
\label{equ:H1}
\end{equation}
where the sum runs over all the nearest-neighbor spin pairs 
$\langle ij \rangle$.
Each spin, \mbox{\boldmath $S$}, is defined by angle, 
$\theta_i$, in the XY plane:
$\mbox{\boldmath $S$}_i = (\cos\theta_i,\sin\theta_i)$.
The angle $\theta _i$ takes continuous values in $[0,2\pi)$, but we divide
it to 1024 discrete states in this paper.
Effects of the discreteness are checked to be negligible by comparing data
of 1024-state simulations with those of 2048-state simulations.
The interactions $J_{ij}$ take two values of $+J$ and $-J$
with the same probability.
The temperature $T$ is scaled by $J$.
Linear lattice sizes are denoted by $L$. 
Total number of spins is $N = L \times L \times (L+1)$, and
skewed periodic boundary conditions are imposed.
Spins are updated by the single-spin-flip algorithm
using the conventional Metropolis probability.

Physical quantities observed in our simulations are 
the spin-glass susceptibility $\chi_\mathrm{sg}$,
the chiral-glass susceptibility $\chi_\mathrm{cg}$,
the Binder parameters $g_\mathrm{sg}$ and $g_\mathrm{cg}$,
the glass correlation length $\xi_\mathrm{sg}$ and $\xi_\mathrm{cg}$.
The spin-glass susceptibility is defined by
the following expression.
\begin{equation}
\chi_\mathrm{sg}\equiv\frac{1}{N}
\left[
\sum_{i,j}\langle \mbox{\boldmath $S$}_i \cdot \mbox{\boldmath $S$}_j \rangle^2
\right]
= 
\frac{2N}{m(m-1)}
\left[
\sum_{A > B, \mu, \nu} (q_{\mu, \nu}^{A B})^2
\right]
\label{equ:xsg}
\end{equation}
The subscripts $\mu$ and $\nu$ denote two components, $x$ and $y$,
of spin variables $\mbox{\boldmath $S$}$.
The thermal average is denoted by $\langle \cdots \rangle$, and 
the random-bond configurational average is denoted by $[ \cdots ]$.
The thermal average is replaced by an average over independent real
replicas:  $\langle \cdots \rangle \to (1/m)\sum_A^m$.
Replica number is denoted by $m$.
Superscripts $A$ and $B$ are the replica indices.
The number  $m$ determines resolution of the thermal average.
It should be large in order to improve the accuracy of average.
We prepare 32 replicas ($m=32$) with different initial spin configurations
in this paper.
Each replica is updated in parallel with a different random number sequence.
An overlap between two replicas is defined by
\begin{equation}
  q_{\mu \nu}^{A B} \equiv \frac{1}{N}
   \sum _i {S}_{i \mu}^{(A)} {S}_{i \nu}^{(B)} .
\label{equ:qsg}
\end{equation}
The chiral-glass susceptibility\cite{KawamuraXY2} is similarly
defined by 
\begin{equation}
  \chi_\mathrm{cg} \equiv \frac{6N}{m(m-1)}
\left[
\sum_{A > B}
(q_{\kappa}^{A B})^2
\right] ,
\label{equ:xcg}
\end{equation}
where
\begin{eqnarray}
  q_{\kappa}^{A B}&\equiv&\frac{1}{3N}\sum _{\alpha} \,
     {\kappa}_{\alpha}^{(A)} {\kappa}_{\alpha}^{(B)},
\label{equ:qcg}
\\
 \kappa _{\alpha}^{(A)}&\equiv&\frac{1}{2 \sqrt{2}} 
 \sum_{\langle jk\rangle\in\alpha} J_{jk} 
 \sin{( \theta _j^{(A)} - \theta _k^{(A)} )}.
\label{equ:localk}
\end{eqnarray}
A local chirality variable, $\kappa_{\alpha}^{(A)}$, is defined at each square
plaquette $\alpha$ that consists of the nearest-neighbor bonds
$\langle jk\rangle$. 

The Binder parameters for the spin glass and for the chiral glass
are calculated through the following expressions.
Here, we have also replaced the thermal averages by the average over 
real replicas.
\begin{eqnarray}
g_\mathrm{sg}&=&3-2
\left(
\frac
{\sum\limits_{\mu, \nu, \delta, \rho}
 \left[
    \frac{2}{m(m-1)}\sum\limits_{A > B}^m 
    (q_{\mu \nu}^{AB})^2 (q_{\delta \rho}^{AB})^2
\right]}
{\sum\limits_{\mu, \nu}
 \left[
     \frac{2}{m(m-1)}\sum\limits_{A > B}^m 
(q_{\mu \nu}^{AB})^2
\right]}
\right) 
\label{equ:binder}
\\
g_\mathrm{cg}&=&\frac{1}{2}
\left(
3-
\frac{ \left[
    \frac{2}{m(m-1)}\sum\limits_{A > B}^m 
    (q_{\kappa}^{AB})^4
\right]}{
 \left[
     \frac{2}{m(m-1)}\sum\limits_{A > B}^m 
(q_{\kappa}^{AB})^2
\right]}
\right)
\label{equ:binderc}
\end  {eqnarray}
Since the Binder parameters lose size dependences at the transition temperature,
the nonequilibrium relaxation functions exhibit algebraic
divergence with an exponent $d/z$.\cite{Blundell} 
From the exponent we can obtain a value of the dynamic exponent $z$.
It also serves as another check of the spin-glass transition.
If the susceptibility and the Binder parameter exhibit algebraic divergence,
it is very certain that the phase transition occurs.
Kawamura and Li \cite{KawamuraXY3} argued that the Binder parameter of the 
chiral glass shows a negative dip at the transition temperature.
They observed crossing behavior of the Binder parameter on the negative side.
We observe nonequilibrium relaxation of the Binder parameter in order to check
this negative crossing: it should diverge algebraically on the negative side.

A spin-glass correlation function is defined by the following expression.
\begin{eqnarray}
f_\mathrm{sg}(l) &\equiv& \left[\sum_i^N\langle
 \mbox{\boldmath $S$}_i \cdot \mbox{\boldmath $S$}_{i+l} \rangle^2\right]
\label{eq:sgcordef}
\\
&=& \left[ 
\frac{2}{m(m-1)}
\sum_{A>B, i, \mu\nu}
q_{\mu\nu}^{AB}(i)
q_{\mu\nu}^{AB}(i+l)
\right]
\nonumber
\end{eqnarray}
A length between two spins is denoted by $l$.
Here,
$
q_{\mu\nu}^{AB}(i)=
S_{i,\mu}^{(A)}S_{i,\nu}^{(B)}
$
is a spin overlap on the $i$ site.
A chiral-glass correlation function is defined by the following.
\begin{equation}
f_\mathrm{cg}(l) = \left[ 
\frac{2}{3m(m-1)}\sum_{A>B}
\sum_{\alpha}
q_{\kappa}^{AB}(\alpha)
q_{\kappa}^{AB}(\alpha+l)
\right]
\end{equation}
Here,
$
q_{\kappa}^{AB}(\alpha)=
\kappa_{\alpha}^{(A)}\kappa_{\alpha}^{(B)}
$
is a local overlap of the chirality.
We only consider a distance $l$ between two parallel square plaquettes.
We estimate a spin-glass correlation length $\xi_\mathrm{sg}$
 and a chiral-glass correlation length $\xi_\mathrm{cg}$ fitting the correlation
functions by the following expression.
\begin{equation}
f_\mathrm{sg/cg}(l)  \sim  \frac{1}{l^{d-\gamma/\nu}}
\exp\left( -\frac{l}{\xi_\mathrm{sg/cg}}\right)
\label{eq:fsg}
\end{equation}
Exponent $d$ is a dimension of the lattice ($d=3$),  $\gamma$
is the glass susceptibility exponent and $\nu$ is the glass correlation length
exponent.

\subsection{Nonequilibrium relaxation method}

We start simulations from random spin configurations.
This is a paramagnetic state at $T=\infty$.
The temperature is quenched to a finite value from the first Monte Carlo step.
At each step we observe physical quantities and obtain the relaxation functions.
Changing an initial spin state, a random bond configuration, 
and a random number sequence we start
another simulation and obtain another set of the relaxation functions.
We take averages of relaxation functions over these independent Monte 
Carlo simulations.
The relaxation functions of the susceptibility and the Binder parameter
multiplied by $N$ exhibit algebraic divergence at the transition temperature:
$\chi_\mathrm{sg/cg} \propto t^{\gamma/z\nu}$ and 
$g_\mathrm{sg/cg}\times N \propto t^{d/z}$.
A relaxation function of the correlation length also diverges algebraically
as $\xi_\mathrm{sg/cg}\propto t^{1/z}$.
We try to find such a temperature first observing raw relaxation functions.
If a temperature is higher than the transition temperature, the susceptibility
converges to a finite value even in the infinite-size system.
The lowest temperature at which the relaxation data exhibit the converging
behavior is the upper bound for the transition temperature.

The second method to estimate the transition temperature is the
finite-time scaling analysis \cite{OzIto2}.
This is a direct interpretation of the conventional 
finite-size scaling analysis by considering the dynamic scaling 
hypothesis: $\tau \sim \xi^z$.
A term ``size'' is replaced by a term ``time'' by this relation.
We collect relaxation functions of the spin-glass susceptibility at various
temperatures.
They should fall onto a single universal curve if we properly choose the
transition temperature, $T_\mathrm{sg}$, and critical exponents, $\gamma$ and
$z\nu$.
The following equation is the scaling formula.
\begin{equation}
\chi_\mathrm{sg}(t) t^{-\gamma/z \nu} =  
  \tilde{\chi}_{\text{sg}}(t/\tau(T)), ~~ \tau(T)\propto 
(T-T_\mathrm{sg})^{-z\nu}
\label{equ:xsg_scl}
\end{equation}
Here, $\tilde{\chi}$ denotes a universal scaling function.
Monte Carlo step $t$ is scaled by correlation time $\tau(T)$ which diverges
algebraically at the transition temperature.
We plot data of the spin-glass susceptibility multiplied by $t^{-\gamma/z\nu}$
against $t/(T-T_\mathrm{sg})^{-z\nu}$.
An estimation of the chiral-glass transition temperature is performed by
the same procedure.

At the estimated transition temperature we observe relaxation functions of 
the Binder parameter and the correlation length.
They should exhibit algebraic divergence.
The observation supports the occurrence of the transition.
The dynamic exponent $z$ is estimated by these relaxation functions.
The exponent of the Binder parameter is $d/z$ and that of the correlation
length is $1/z$.
We can check consistency of the estimates comparing these two values.

\subsection{New criterion for scaling analysis}

Finite-size/time-scaling analysis is a powerful tool to investigate
critical phenomena.
The method has been applied to various systems successfully.
However, it sometimes has ambiguity obtaining correct values of
the transition temperature and the critical exponents.
One example is a spin-glass transition problem in the $\pm J$ Heisenberg model
in three dimensions.
Olive {\it et al.} \cite{Olive}
performed a finite-size-scaling analysis of
the spin-glass susceptibility and concluded that the phase transition does
not occur.
Matsubara {\it et al.}\cite{Matsubara4} reported in the preprint version
that the same analysis may give a finite transition
temperature.
A difference of these two analyses is a temperature range of data used
in the scaling.
We also experience this kind of ambiguity in the present paper.

Determination of good scaling is usually done by the looks of scaling plots.
All the scaled data are fitted by some polynomial functions.
One which makes the least standard deviation is selected to be the
best scaling result.
It may change if we use a different set of
temperature/size/time ranges in the analysis.
This change is systematic.

For example, we consider a situation when a ratio of exponents 
$\gamma/z\nu$ in Eq. (\ref{equ:xsg_scl}) is larger than the correct value.
The left-hand-side of the scaling formula becomes smaller and smaller
as time increases.
This tendency is stronger at lower temperatures where the correlation time
is relatively large.
In order to make scaled data fall onto a single curve we must choose 
$\tau(T)$ at lower temperatures to be smaller than the correct values.
This is because the universal scaling function $\tilde{\chi}$ is a
decreasing function with respect to $t/\tau(T)$.
Since $\tau(T)\propto (T-T_\mathrm{sg})^{-z\nu}$, the transition temperature 
is underestimated when a ratio $\gamma/z\nu$ is larger than the correct value.
As we discard data at higher temperatures and restrict our analysis to
lower temperature ranges,
the transition temperature is more and more underestimated.
Contrarily, the transition temperature is more and more overestimated
when the ratio $\gamma/z\nu$ is smaller than the correct value.
When the ratio is correct, the correct transition temperature is obtained 
and the value is robust against changes of the temperature range.

Our new criterion for the scaling is to find a temperature and critical 
exponents so that the obtained results are independent from the
temperature range of data used in the scaling analysis.
First, we set one temperature range.
We obtain the transition temperature
which gives the best scaling plot at each point of $\gamma/z\nu$.
The obtained transition temperature is plotted against $\gamma/z\nu$.
Changing the temperature range we perform the same analysis and obtain 
another plot for the transition temperature.
These plots exhibit systematic behavior as mentioned above.
They cross at one point, which is the most probable estimate for the
transition temperature.

\begin{figure}
  \begin{center}
  \resizebox{0.45\textwidth}{!}{\includegraphics{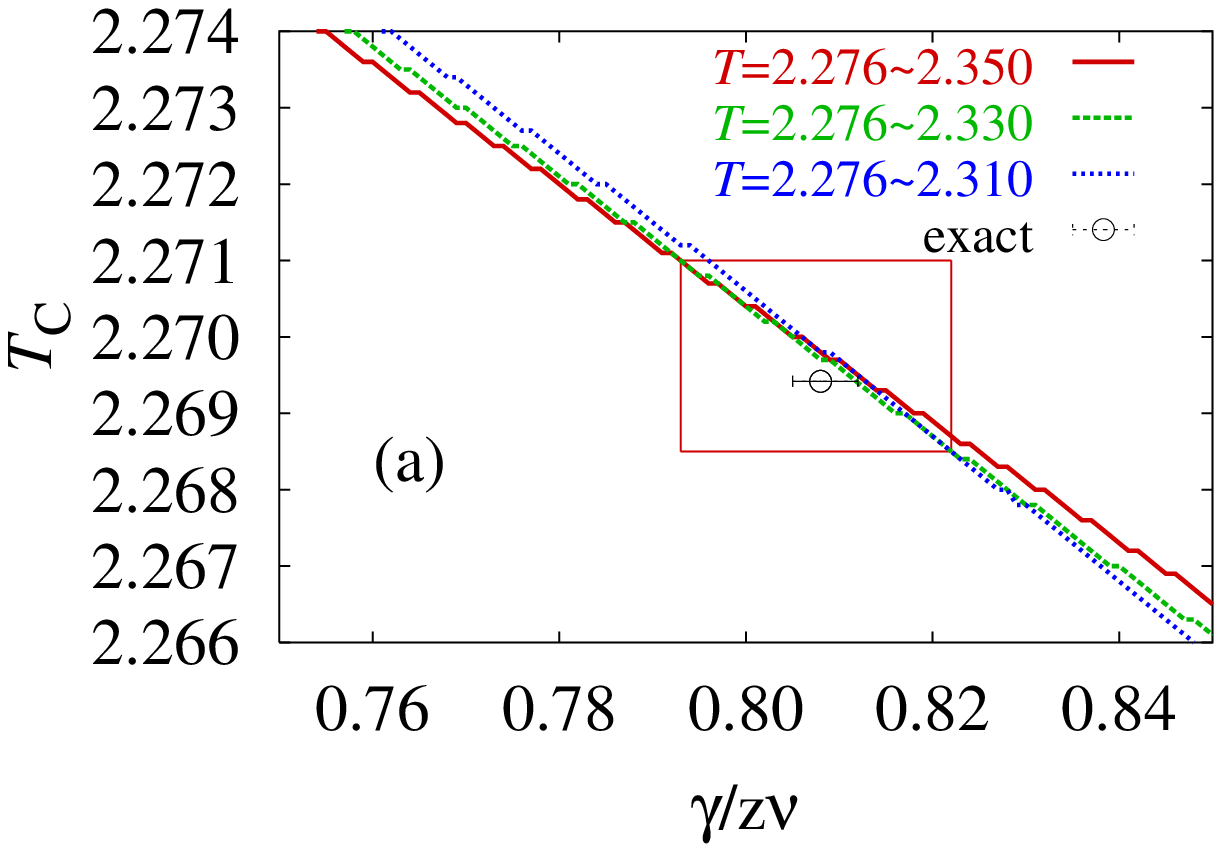}}
  \resizebox{0.45\textwidth}{!}{\includegraphics{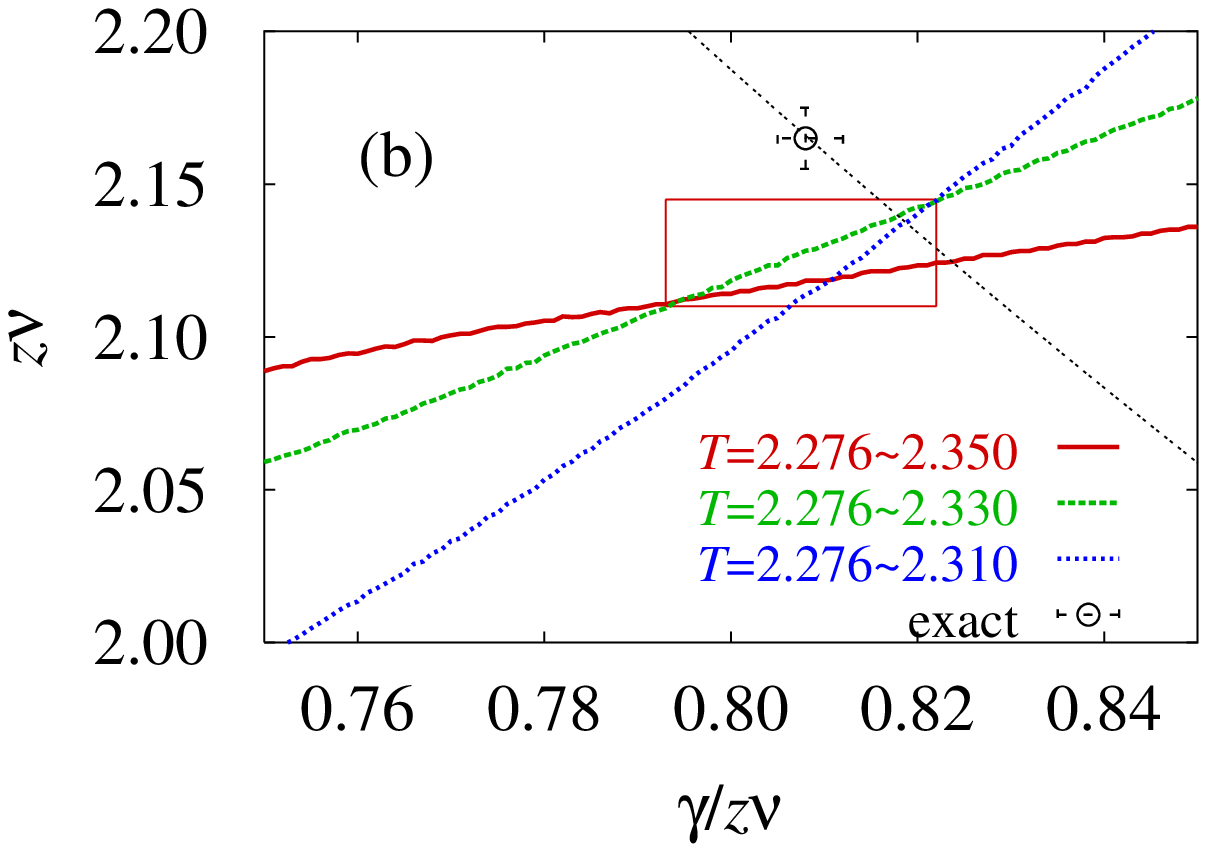}}
  \end{center}
  \caption{
Results of the finite-time-scaling analysis with our new criterion
in a ferromagnetic Ising model in two dimensions.
An exact estimate using $z=2.165(10)$ \cite{2dising}
 is depicted by a symbol. 
Boxes depict error bars.
(a) $T_\mathrm{c}$ versus $\gamma/z\nu$. 
(b) $z\nu$ versus $\gamma/z\nu$.
A thin line is an exact line regardless a value of $z$.
}
\label{fig:ising}
\end{figure}

Figure \ref{fig:ising} shows an example of the analysis with the new scaling
criterion in the ferromagnetic Ising model in two dimensions.
The lattice size is $499\times 500$ with skewed periodic boundary conditions.
The Monte Carlo step is 2400.
The first 300 steps are discarded, which contain initial relaxation.
Simulations are performed at nineteen different temperatures ranging
$2.276 \le T \le 2.35$.
Averages over 124000 independent Monte Carlo runs are taken.
When a ratio $\gamma/z\nu$ is larger,
estimates of the transition temperature go down 
as the temperature range shrinks to lower temperatures.
They go up when a ratio is smaller.
The plots cross near the exact point.
Deviation from the exact value is larger for
the exponent $z\nu$ as shown in Fig. \ref{fig:ising}(b).
Crossing behavior is rough in this case.
An estimation of the exponent is difficult compared to the
transition temperature.
We perform this scaling procedure to estimate the spin-glass transition
temperature and the chiral-glass transition temperature.

\section{Results}
\label{sec:results}

\subsection{Relaxation of the susceptibility}
First, we observe raw relaxation functions of the glass susceptibility.
We must check time scale when the finite-size effects appear in the
relaxation functions.
Data must be free from the finite-size effects
because the nonequilibrium relaxation method is based upon 
taking the infinite-size limit before the equilibrium limit is taken.
The susceptibility data influenced  by the size effects exhibit converging
behavior even though it should diverge in the infinite-size limit.
It misleads us into thinking that the temperature is in the paramagnetic phase.

\begin{figure}
  \begin{center}
  \resizebox{0.45\textwidth}{!}{\includegraphics{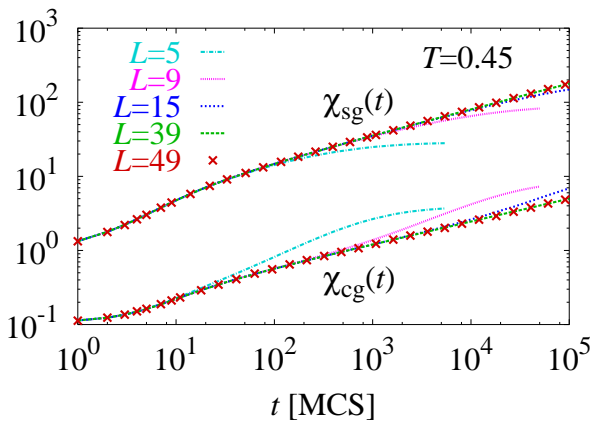}}
  \end{center}
  \caption{
Finite-size effects of relaxation functions
of $\chi_\mathrm{sg}(t)$ and $\chi_\mathrm{cg}(t)$.
The temperature $T=0.45$ is near our estimate for the 
transition temperature.
There is no size dependence for $L=39$ within $10^5$ Monte Carlo step.
}
\label{fig:size1}
\end{figure}

Figure \ref{fig:size1} shows the appearance of the finite-size effects of 
$\chi_\mathrm{sg}(t)$ and $\chi_\mathrm{cg}(t)$.
When a size is small, a nonequilibrium relaxation function deviates from 
the size-independent relaxation curve at a characteristic time.
The system finds itself finite at this time.
It behaves as an infinite-size system before the time.
There is no difference between data of $L=39$ and those of $L=49$.
Therefore, relaxation functions of $L=39$ within $t\le 10^5$ are regarded as 
of the infinite-size system.
We mainly use this set of size and time limits in this paper.
Number of random bond configurations is typically more than several hundreds.
It varies depending on system sizes, Monte Carlo steps, and physical quantities.
For example, quite a few configurations are necessary in order to obtain 
beautiful relaxation functions for the Binder parameter
and the correlation length.

It is also noted in Fig. \ref{fig:size1}
that behavior of the finite-size effects of the chiral-glass
susceptibility is opposite to the spin-glass one.
The spin-glass susceptibility deviates from the diverging curve 
just to converge to a finite value.
It is natural behavior.
On the other hand, the chiral-glass susceptibility increases
when the size effects appear. 
It eventually converges to a finite value and is overtaken 
by the infinite-size one.
This strange behavior is also observed in the $\pm J$ Heisenberg model in 
three dimensions.\cite{Nakamura}

\begin{figure}
  \begin{center}
  \resizebox{0.45\textwidth}{!}{\includegraphics{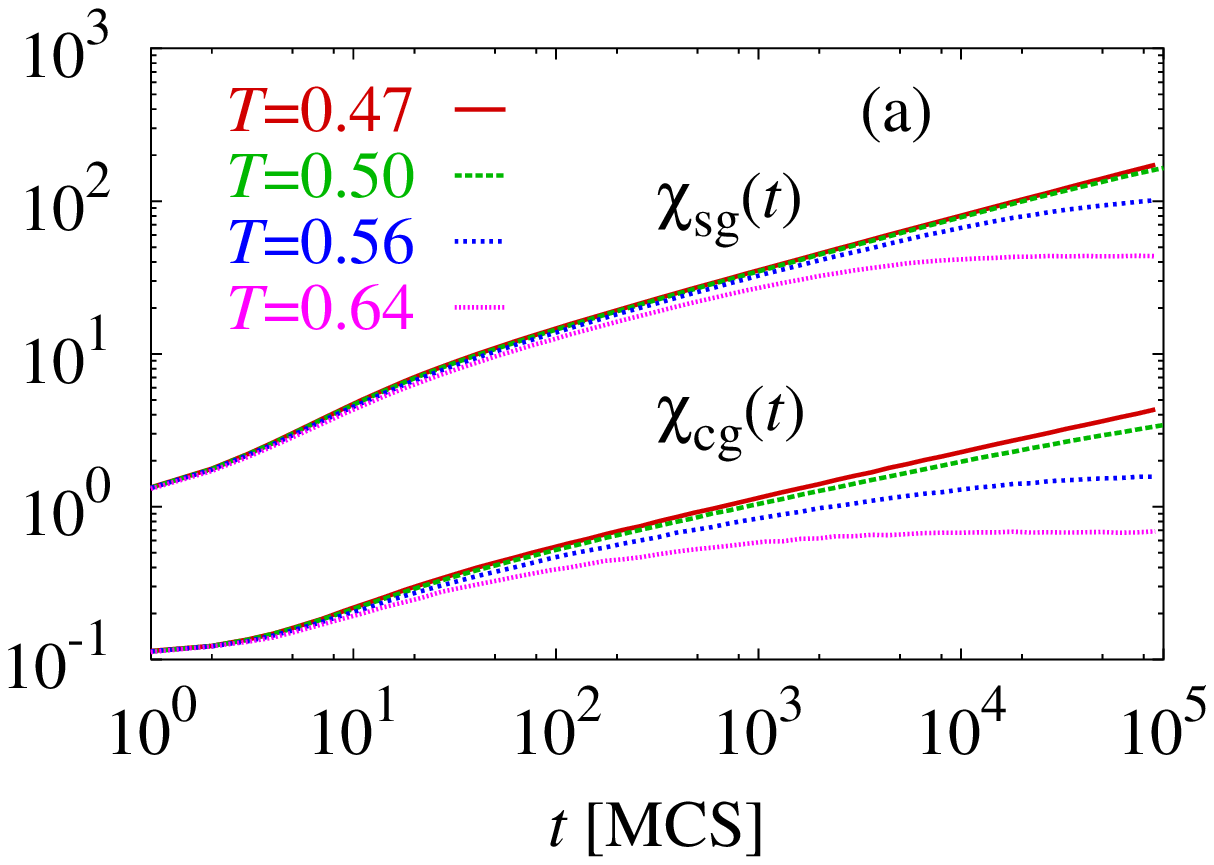}}
  \resizebox{0.45\textwidth}{!}{\includegraphics{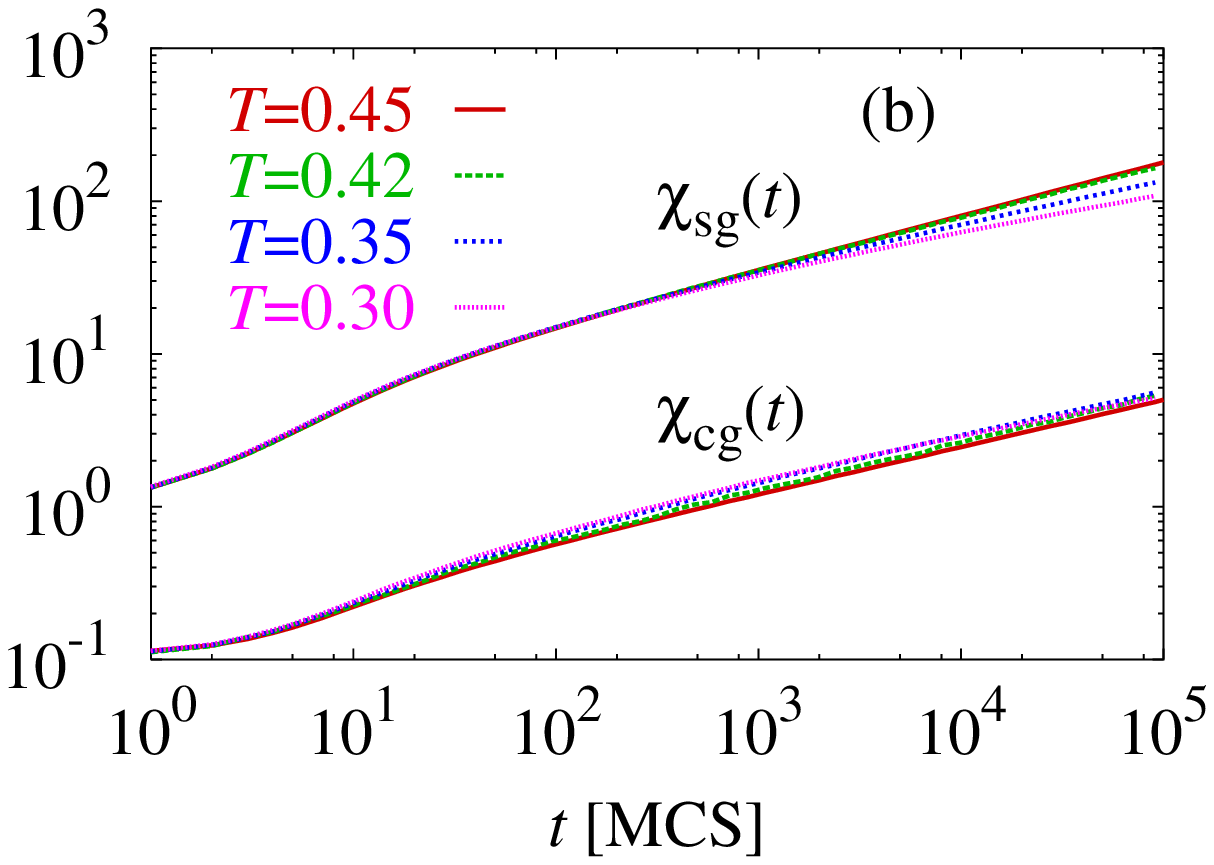}}
  \resizebox{0.45\textwidth}{!}{\includegraphics{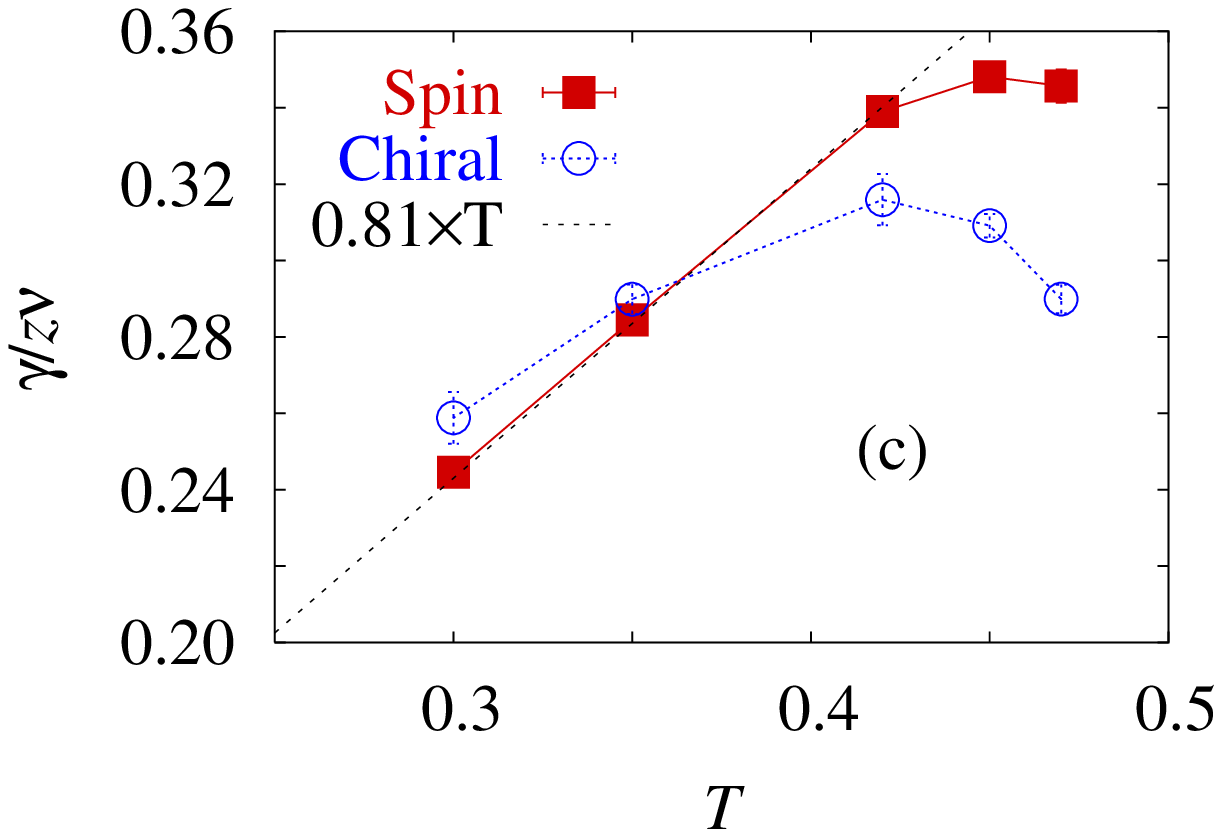}}
  \end{center}
  \caption{
Relaxation functions of $\chi_\mathrm{sg}$ and $\chi_\mathrm{cg}$ at various
temperatures.
(a) Plots at $T \ge 0.47$: high temperature side.
(b) Plots at $T \le 0.45$: low temperature side.
(c) Exponents of algebraic divergence in $\chi_\mathrm{sg}$ and 
$\chi_\mathrm{cg}$ at lower temperatures.
}
\label{fig:chi1}
\end{figure}

Figure \ref{fig:chi1}(a) shows relaxation functions of 
 $\chi_\mathrm{sg}(t)$ and $\chi_\mathrm{cg}(t)$ for  $T \ge 0.47$.
This is a plot on the high temperature side.
It is clear that there is no spin-glass/chiral-glass transition at $T > 0.50$.
Both susceptibility exhibit converging behavior.
They exhibit algebraic divergence at $T = 0.47$.
It suggests that the spin-glass transition occurs
at a temperature near $T=0.47$.
Figure \ref{fig:chi1}(b) shows a plot at lower temperatures.
The spin-glass susceptibility diverges algebraically with exponents decreasing
with the temperatures.
The exponent is largest at $T=0.45$.
We consider it to be near the transition temperature.
The same behavior is observed in the $\pm J$ Ising model in 
three dimensions below the spin-glass transition 
temperature.\cite{Komori,totaIsing}
Slow dynamics appear in the low-temperature spin-glass phase.
Relaxation of the glass susceptibility may become algebraically slow.
The chiral-glass susceptibility also exhibit algebraic divergence
at lower temperatures.
The amplitude as well as the exponent is dependent on the temperatures.

Temperature dependences of the exponents are shown in Fig.~\ref{fig:chi1}(c).
Algebraic divergence starts appearing at $T\sim 0.47$.
An exponent of the spin-glass susceptibility takes a maximum value 
near $T=0.45$.
It is considered in the vicinity of the transition temperature.
The exponent decreases as the temperature decreases.
This temperature dependence is linear with $T$:
\begin{equation}
\gamma/z(T)\nu = 0.81T,~~ 1/z(T)\sim 0.4T.
\label{eq:zofT}
\end{equation}
Here, we have used our final estimate for a ratio $\gamma/\nu=2.0(2)$ obtained
in Sec. \ref{sec:NERcorlen}.
The temperature dependence of $1/z$ in the Ising model with Gaussian 
bond distributions is reported by Komori {\it et al.} \cite{Komori} to be
$1/z(T)=0.17T$.
The low-temperature phase in the present XY model is
similar to that of the Ising model.

Temperature dependences of an exponent of the chiral-glass susceptibility
is strange compared with the spin-glass one.
The algebraic divergence begins at $T=0.47$.
As the temperature decreases, the exponent first increases and takes a
maximum value at $T=0.42$ and then decreases with temperature.
The temperature of the maximum exponent is lower 
than that of the spin-glass one.

From the relaxation functions of the susceptibility
it is very possible that the spin-glass phase transition occurs at 
$T=0.45\pm 0.03$.
The chiral-glass transition also occurs within this temperature range.
Since the chirality trivially freezes if the spin freezes,
these two transitions may occur simultaneously.
The low-temperature relaxation functions of the spin-glass susceptibility
exhibit a property of the spin-glass phase:
an inverse of the dynamic exponent is linearly dependent on the temperature.
We check these findings performing the finite-time scaling analysis.
Stress is put upon consistency of a transition temperature and a ratio of
critical exponents $\gamma/z\nu$ at that temperature.

\subsection{Finite-time-scaling analyses}
\begin{figure}
  \begin{center}
  \resizebox{0.45\textwidth}{!}{\includegraphics{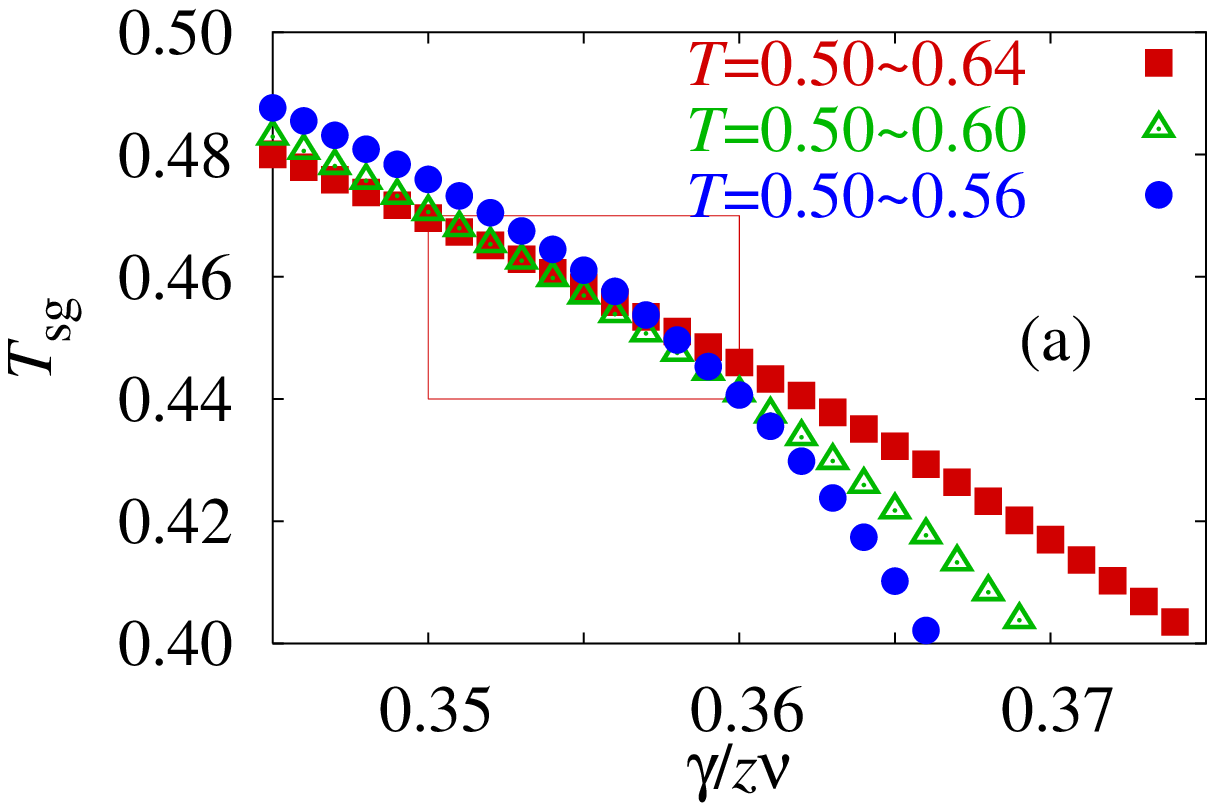}}
  \resizebox{0.45\textwidth}{!}{\includegraphics{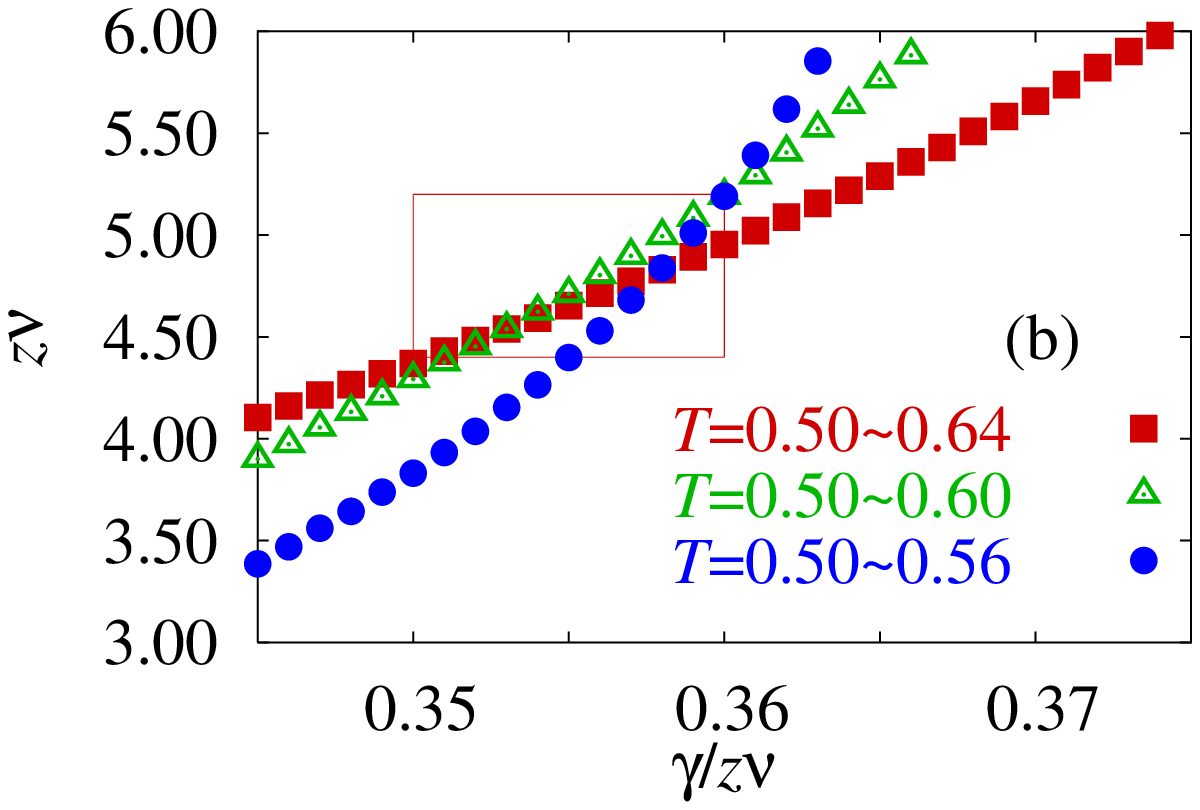}}
  \resizebox{0.45\textwidth}{!}{\includegraphics{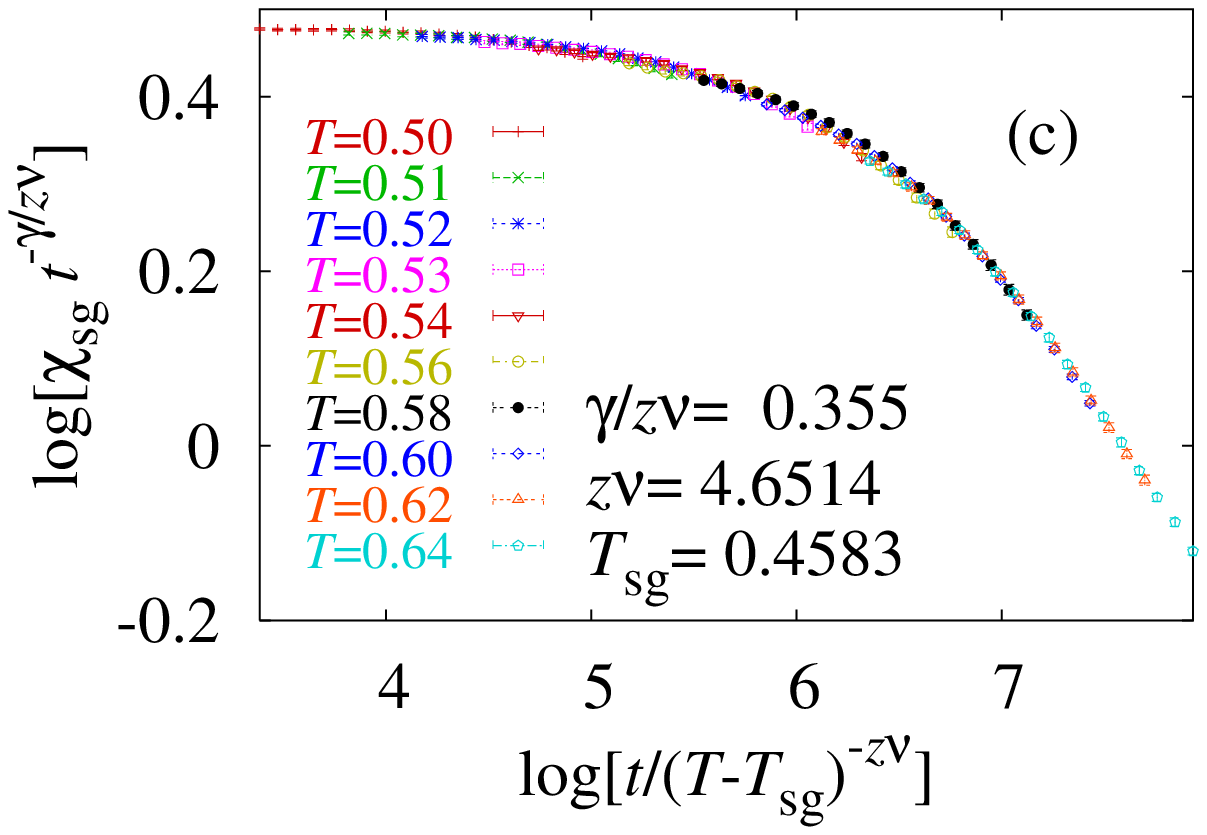}}
  \end{center}
  \caption{
Results of the present finite-time-scaling analysis on the spin-glass
transition.
(a) $T_\mathrm{sg}$ versus $\gamma/z\nu$.
(b) $z\nu$ versus $\gamma/z\nu$.
(c) A scaling plot using the obtained parameters.
}
\label{fig:scl1}
\end{figure}

We perform a finite-time-scaling analysis with our new criterion.
Figures \ref{fig:scl1}(a) and \ref{fig:scl1}(b) show results of the scaling 
on the spin-glass transition.
For each value of $\gamma/z\nu$ we choose $T_\mathrm{sg}$ and $z\nu$ so that
fitting error of the scaled data becomes smallest. 
The fitting is done by a polynomial expression up to the 7th order.
Relaxation data of the first two thousand steps are discarded.
We considered them to be the initial relaxation.
The scaling results are dependent on the temperature range.
The dependence is also as described in Sec. \ref{sec:method}:
when $\gamma/z\nu$ is larger, the transition temperature becomes underestimated
as high temperature data are discarded, and vice versa.
From the crossing point we obtain our estimates as
\begin{eqnarray}
T_\mathrm{sg} &=& 0.455 \pm 0.015\\
\gamma/z\nu&=& 0.355 \pm 0.005\\
z\nu &=& 4.8 \pm 0.4.
\end{eqnarray}

Figure \ref{fig:scl1}(c) shows a scaling plot using the obtained results.
All data fall onto a single scaling function very nicely.
The obtained transition temperature is higher than other estimates.
Granato \cite{Granato} gave $T_\mathrm{sg}=0.39(2)$, $\nu=1.2(2)$ and
$z=4.4(3)$.
A value of $z\nu$ is consistent with our value.
Lee and Young \cite{Lee} gave for the Gaussian model 
$T_\mathrm{sg}=0.34(2)$ and $\nu=1.2(2)$.
Our estimate for $\nu$ is $\nu=0.85(15)$, where we use a value of $z$ 
which will be obtained in Sec. \ref{sec:NERcorlen} (Eq. (\ref{eq:zfinal})).
Though the transition temperature is quite different, 
critical exponents are consistent.

Values of $T_\mathrm{sg}$ and $\gamma/z\nu$ obtained by the scaling analysis
agree with the relaxation function
of the susceptibility in the previous subsection.
As shown in Fig. \ref{fig:chi1}(c) an exponent of the spin-glass susceptibility
takes a maximum value at around $T=0.45$ and the exponent is 
$\gamma/z\nu=0.348$.
Both values are consistent with the scaling results.
A scaling estimation of $T_\mathrm{sg}=0.4$ is possible
if we artificially choose $\gamma/z\nu$ and the
temperature range of the data used in the scaling analysis.
An example is a set of choices: $\gamma/z\nu=0.366$ and $T=0.50\sim 0.56$
as shown in Fig. \ref{fig:scl1}(a) with circles.
However, it is inconsistent with the relaxation functions of the spin-glass 
susceptibility.
It does not diverge with an exponent $\gamma/z\nu=0.366$ at $T=0.4$.
Only the scaling results that are independent from the
temperature ranges (a crossing point in the figure) 
become consistent with the raw relaxation functions.

\begin{figure}
  \begin{center}
  \resizebox{0.45\textwidth}{!}{\includegraphics{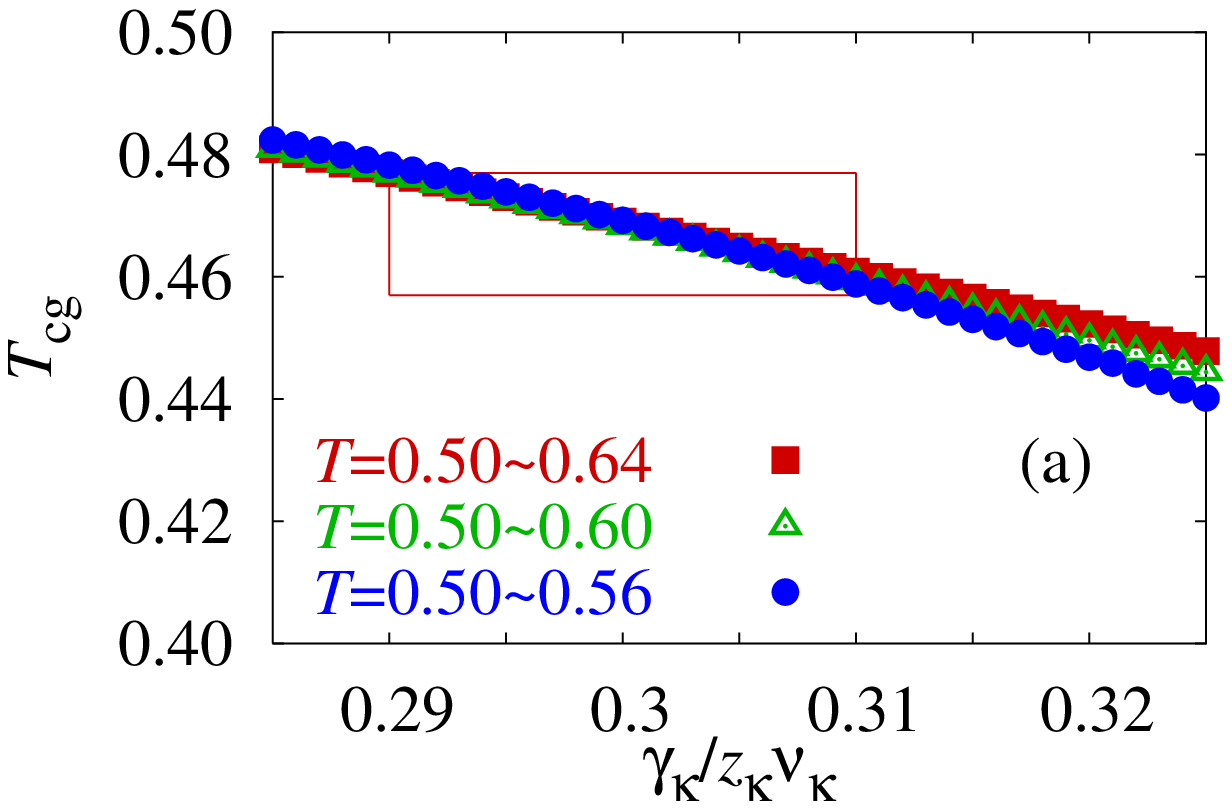}}
  \resizebox{0.45\textwidth}{!}{\includegraphics{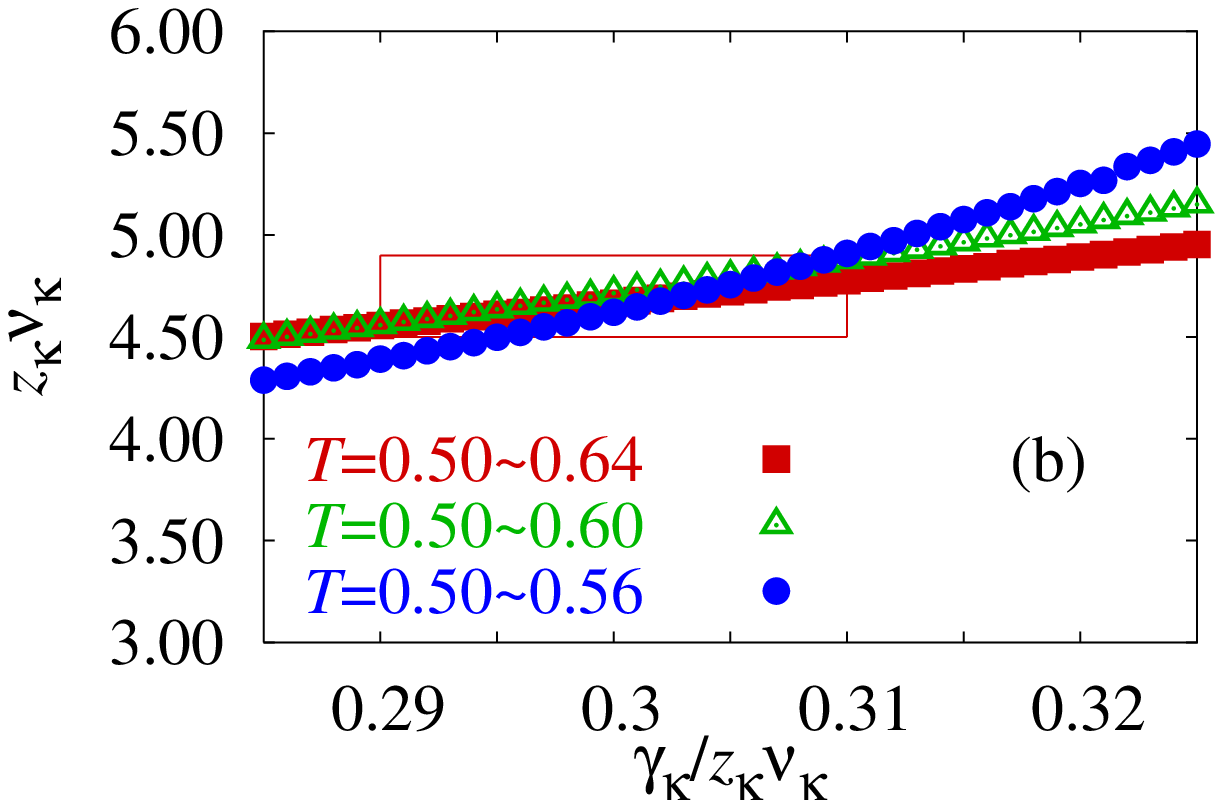}}
  \resizebox{0.45\textwidth}{!}{\includegraphics{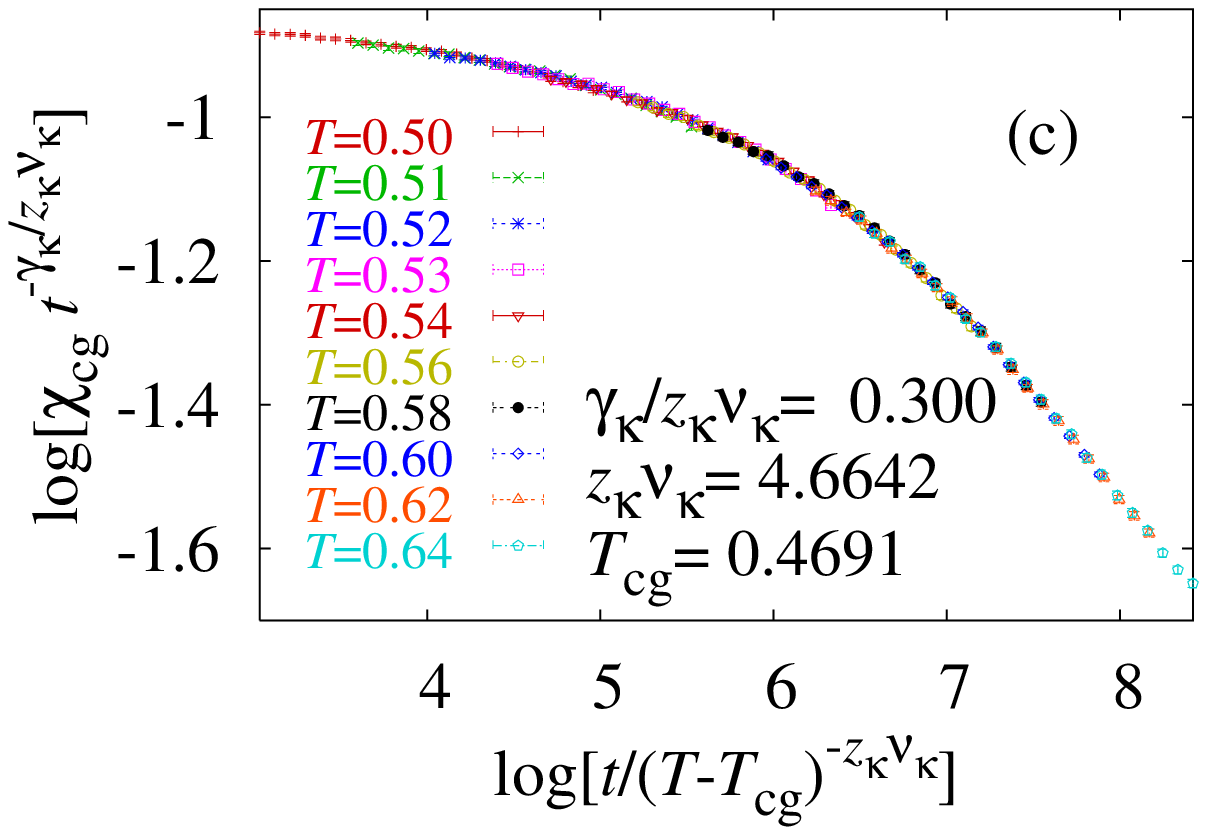}}
  \end{center}
  \caption{
Results of the present finite-time-scaling analysis on the chiral-glass
transition.
(a) $T_\mathrm{sg}$ versus $\gamma/z\nu$.
(b) $z\nu$ versus $\gamma/z\nu$.
(c) A scaling plot using the obtained parameters.
}
\label{fig:scl2}
\end{figure}

Figures \ref{fig:scl2}(a), \ref{fig:scl2}(b) and \ref{fig:scl2}(c) 
shows results on the chiral-glass transition.
Estimations of the transition temperature and critical exponents are
\begin{eqnarray}
T_\mathrm{cg} &=& 0.467 \pm 0.010\\
\gamma_{\kappa}/z_{\kappa}\nu_{\kappa}&=& 0.30 \pm 0.01\\
z_{\kappa}\nu_{\kappa} &=& 4.7 \pm 0.2.
\end{eqnarray}
Subscripts $\kappa$ of the critical exponents denote chirality.
Kawamura and Li gave $T_\mathrm{cg}=0.39(3)$, $\nu_{\kappa}=1.2(2)$, 
$\gamma_{\kappa}/\nu_{\kappa}=1.85(20)$, and $z_{\kappa}=7.4(10)$.
Our estimation for the transition temperature is higher than their value.
We will obtain a value of $z_{\kappa}=6.3(5)$ from a relaxation function
of the chiral-glass correlation length in Sec. \ref{sec:NERcorlen}.
Then, an exponent $\nu_{\kappa}=0.75(10)$ is obtained.
This value is also smaller than their value. 
However, a ratio of exponents, 
$\gamma_{\kappa}/\nu_{\kappa}=1.9(2)$, is consistent.
As observed in Fig. \ref{fig:chi1}(c) 
an exponent $\gamma_{\kappa}/z_{\kappa}\nu_{\kappa}$ 
of the chiral-glass susceptibility takes a maximum value of 
$\gamma_{\kappa}/z_{\kappa}\nu_{\kappa}=0.315$ at $T=0.42$.
This temperature is consistent with the estimation of Kawamura and Li.
However, this set of $T$ and $\gamma_{\kappa}/z_{\kappa}\nu_{\kappa}$ 
is inconsistent with the scaling results
of Fig. \ref{fig:scl2}(a).
To the contrary, our scaling estimations are consistent with the
relaxation functions of the chiral-glass susceptibility.
We consider that the chiral-glass transition occurs at $T\sim 0.47$.
Algebraic divergence begins at this temperature.
The temperature of the exponent maximum does not correspond to the 
transition temperature of the chirality.

The spin-glass transition temperature and the chiral-glass transition
temperature agree well within the error bars.
Therefore, both transitions are considered to occur simultaneously.
Estimations are consistent with the raw relaxation functions.
The consistency guarantees that our analyses are performed correctly.


\subsection{Relaxation of the Binder parameter}
In order for another check of our scaling results
and for obtaining the dynamic exponent $z$ independently
we have calculated the nonequilibrium relaxation functions of the
Binder parameter at the estimated transition temperature.

\begin{figure}
  \begin{center}
  \resizebox{0.45\textwidth}{!}{\includegraphics{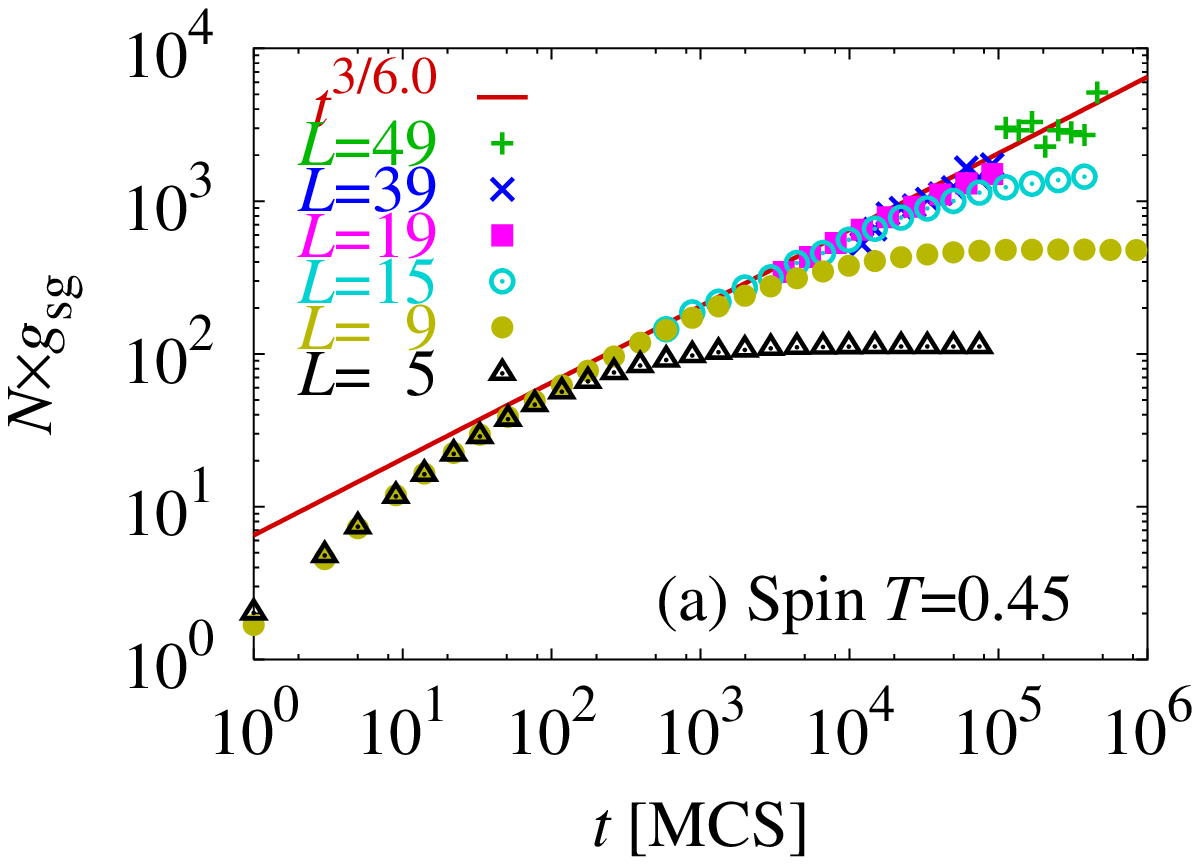}}
  \resizebox{0.45\textwidth}{!}{\includegraphics{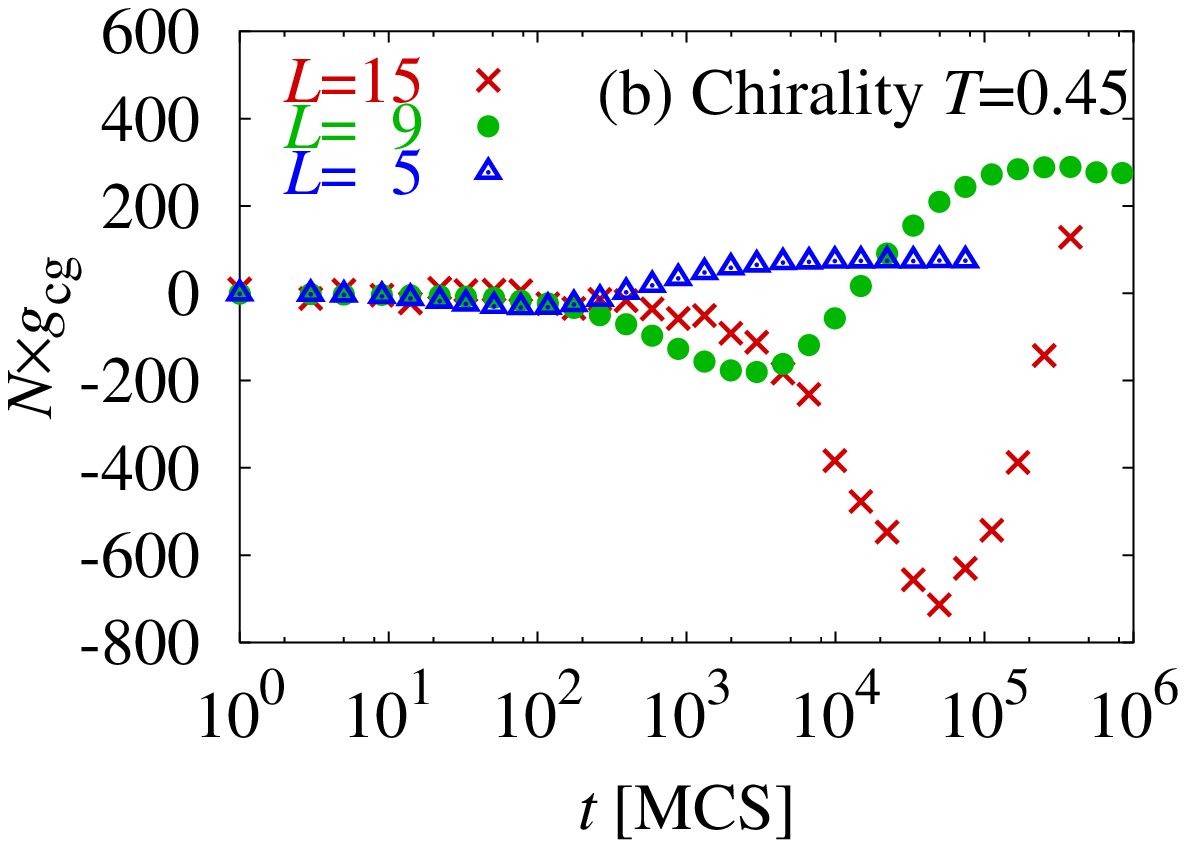}}
  \end{center}
  \caption{
Relaxation functions of the Binder parameters at $T=0.45$.
(a) A log-log plot for the spin-glass one. Algebraic divergence is observed.
(b) A semi-log plot for the chiral-glass one. No divergence is observed.
When the size effects appear, the Binder parameters take negative values
first and converge to positive values.
}
\label{fig:binder1}
\end{figure}

Figure \ref{fig:binder1}(a) shows relaxation functions of the Binder
parameter of the spin, $g_\mathrm{sg}(t)$, for various system sizes.
When a size is small, a relaxation function deviate from the size-independent
relaxation curve. 
Appearance of the finite-size effect is normal as in the case of the
spin-glass susceptibility.
It just converges to a finite value when the size effect appears.
The size-independent curve exhibits an algebraic divergence with an exponent
$d/z$ with $z\sim 6.0\pm 0.5$.
We have also obtained the relaxation functions at $T=0.42$ and $T=0.47$.
(Figures not shown in this paper.)
These temperatures are within the error bar of 
the spin-glass transition temperature.
The algebraic divergence of the Binder parameter is also observed.
It is noted that the dynamic exponent $z$ is systematically 
dependent on the temperature.
The values are $z=6.5\pm 0.7$ at $T=0.42$ and $z=5.7\pm 0.5$ at $T=0.47$.
They are consistent with the temperature dependence of $z$ 
in Eq. (\ref{eq:zofT}): $1/z(T)\sim 0.4T$ within error bars.
The dynamic exponent is estimated to be $ z=6.0 \pm 0.8$
within an error bar of the spin-glass transition temperature.

Relaxation functions of the Binder parameter of the 
chiral-glass shows strange behavior.
The algebraic divergence as in the spin-glass one cannot be observed
at all in Fig. \ref{fig:binder1}(b).
A value of the Binder parameter remains zero 
before the finite-size effects appear.
Then, it decreases and takes negative values.
Taking a minimum value it increases with time and 
eventually converges to a positive value, which is consistent
with a result of Kawamura and Li\cite{KawamuraXY3}.
This strange finite-size effect possibly has the same origin
with that of the chiral-glass susceptibility as shown in Fig. \ref{fig:size1}.
The chiral-glass Binder parameter is considered to take 
zero value at $T>T_\mathrm{cg}$ in the infinite-size system.

\subsection{Relaxation of the correlation length}
\label{sec:NERcorlen}

In this subsection
the final check is made on our analysis concerning the spin-glass transition.
We also answer questions and claims regarding use of 
the nonequilibrium relaxation method in spin-glass problems.

\begin{figure}
  \begin{center}
  \resizebox{0.45\textwidth}{!}{\includegraphics{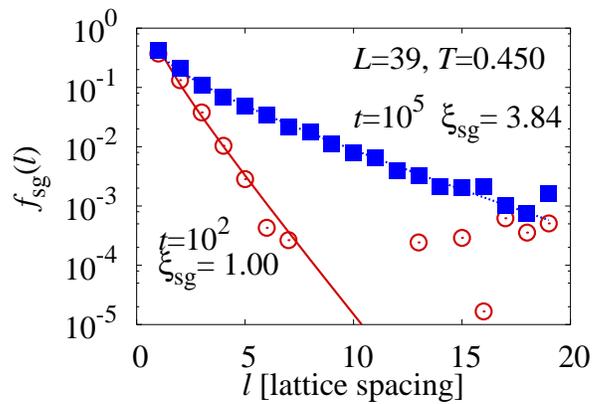}}
  \end{center}
  \caption{
Estimations of a correlation length $\xi_\mathrm{sg}(t)$
by fitting a correlation function $f_\mathrm{sg}(l)$.
Data at $t=10^2$ are depicted by circles and 
those at $t=10^5$ are depicted by squares.
}
\label{fig:corfunc1}
\end{figure}

Figure \ref{fig:corfunc1} shows the spin-glass correlation functions at
Monte Carlo steps $t=10^2$ and $t=10^5$.
The temperature is near the spin-glass transition temperature.
We have taken averages over three directions, $x$, $y$ and $z$, between spins.
The distance is limited within $L/2$ because we have imposed skewed periodic
boundary conditions.
Our simulations start from the perfectly random spin configuration.
There is no correlation between spins and 
the correlation length is zero at $t=0$.
The spin-glass correlation grows with time. 
The correlation decays fast at $t=10^2$ but decay is slow at $t=10^5$.
We estimate the correlation length fitting the correlation functions
by an expression Eq. (\ref{eq:fsg}).
Here, we have set $\gamma/\nu=2.4$.
We obtain a ratio of $\gamma/\nu\simeq 2.1(3)$
from the scaling result of $\gamma/z\nu=0.355(5)$ together with an estimate of
the dynamic exponent $z=6.0(8)$ by the Binder parameter.
We performed the fitting with several values of $\gamma/\nu$ in this range.
Amplitudes of the correlation lengths change with $\gamma/\nu$.
When a value of $\gamma/\nu$ is small, $\xi_\mathrm{sg}$ is overestimated:
it becomes 40\% larger if we set $\gamma/\nu=1.8$.
However, the dynamic-scaling behavior, $\xi\sim t^{1/z}$,
and the exponent $z$ remain the same.
We discard data of the correlation functions which are lower than
$10^{-3}$ in the fitting procedure.
This is resolution of the present simulations.
Numerical data may take negative values below this resolution limit even though
the definition Eq. (\ref{eq:sgcordef}) demands positive values.
The obtained correlation length is not influenced by such garbage data.
The spin-glass correlation length at $t=10^2$ is $\xi_\mathrm{sg}=1.00$, and
$\xi_\mathrm{sg}=3.84$ at $t=10^5$.
The growth is very slow.
We have also obtained the chiral-glass correlation length in the same manner.

\begin{figure}
  \begin{center}
  \resizebox{0.45\textwidth}{!}{\includegraphics{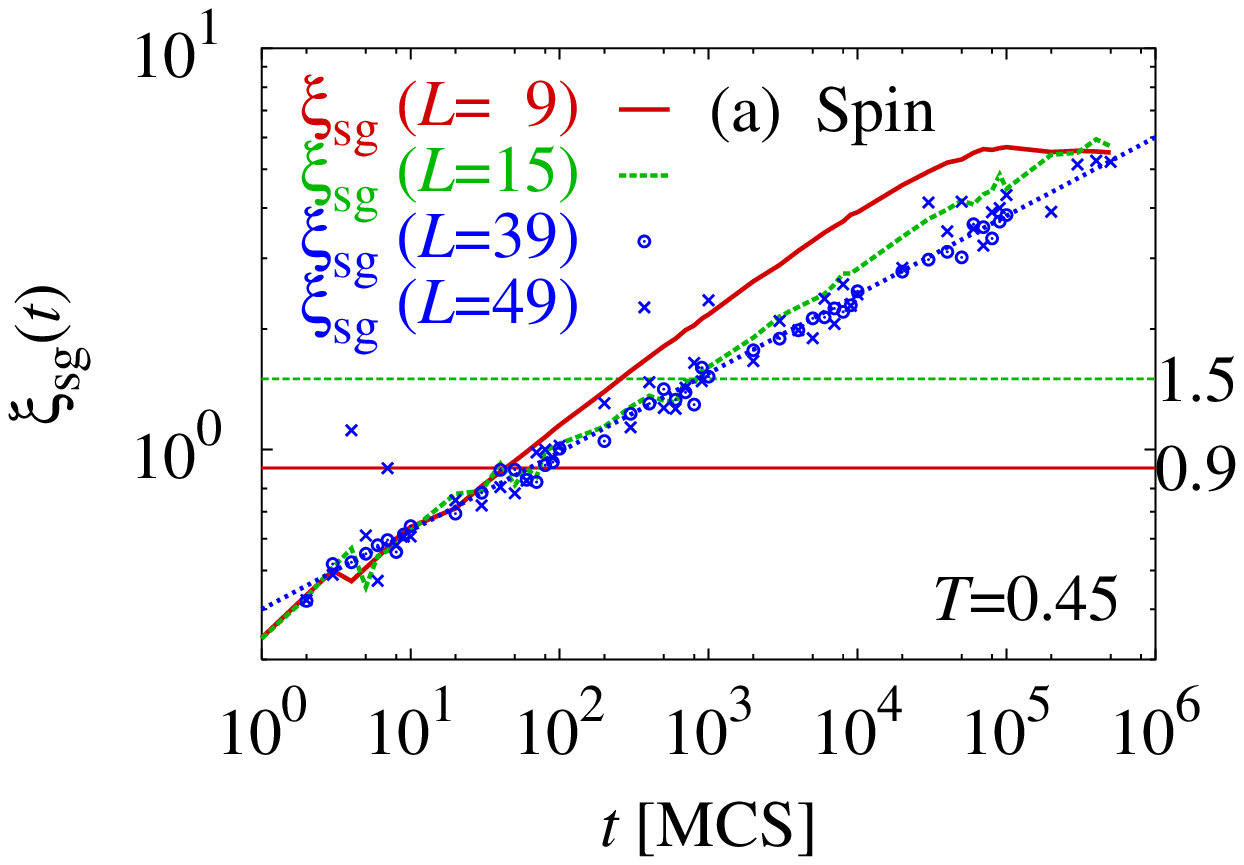}}
  \resizebox{0.45\textwidth}{!}{\includegraphics{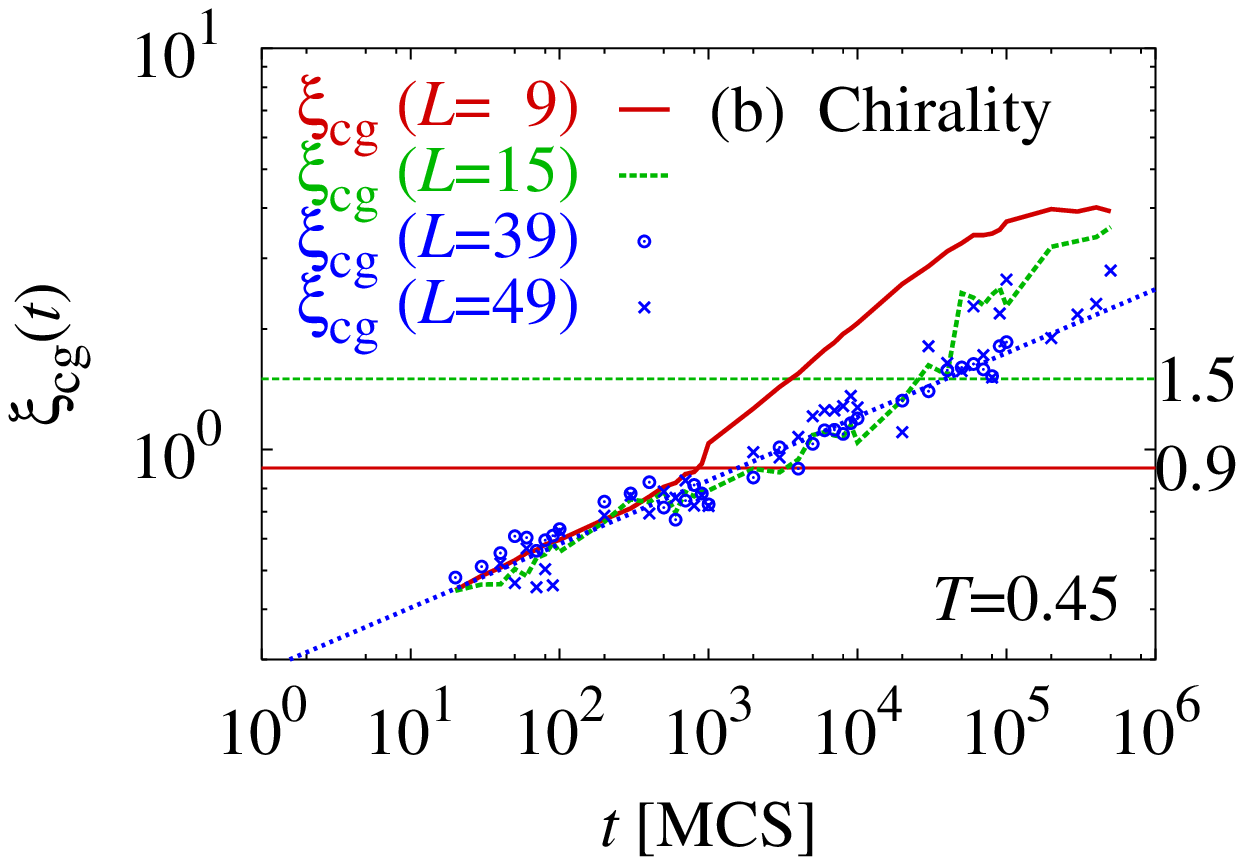}}
  \end{center}
  \caption{
Relaxation functions of 
(a) the spin-glass correlation length and (b) the chiral-glass
correlation length.
Linear sizes of lattices are denoted by $L$.
Fitting lines for the size-independent data are 
$\xi_\mathrm{sg}(t)=0.4t^{1/5.1}$ and
$\xi_\mathrm{cg}(t)=0.3t^{1/6.3}$, respectively.
The size effect appears when the correlation length reaches $L/10$ in 
every case.
}
\label{fig:xi2}
\end{figure}

Figure \ref{fig:xi2} shows relaxation functions of the spin-glass correlation
length and the chiral-glass correlation length.
At each Monte Carlo step we calculate the correlation functions and 
obtain the correlation lengths.
The correlation length exhibits algebraically diverging behavior until
$t=10^6$.
There is no size dependence between data of $L=39$ and those of $L=49$.
We can regard them the infinite-size data.
Random configuration number of the $L=39$ data is 235, and
that of the $L=49$ data is 122.
Deviations from the fitting line are larger for the $L=49$ data.
The dynamic exponent of the spin-glass correlation length is $z=5.1\pm 0.5$.
This value is a little smaller than the estimation by the Binder parameter,
which gave $z=6.0\pm 0.8$.
Both estimates are independent.
Thus, we take an average of them and obtain our final estimate.
\begin{eqnarray}
z&=&5.6 \pm 0.6  ~~~(\mathrm{spin})
\label{eq:zfinal}
\\
z_{\kappa}&=&6.3 \pm 0.5  ~~~(\mathrm{chirality})
\end  {eqnarray}
Using the value of spin and $\gamma/z\nu=0.355(5)$ 
we obtain a ratio of the critical exponents
of the spin-glass susceptibility as $\gamma/\nu=2.0(2)$.

It is remarkable that the spin-glass correlation length starts the diverging
behavior almost from the first step.
A correlation length is about 0.5 lattice spacings.
The chiral-glass correlation length also starts the diverging behavior
at $t=50$ where a correlation length is also 0.5 lattice spacings.
This is a characteristic length above which
we can observe the algebraic divergence of the correlation length.
It is considered to be determined by resolution of simulations.
The figure also shows how the finite-size effects appear in the relaxation
functions of the correlation lengths.
In both plots the finite-size effects appear when the correlation lengths
reach $L/10$.
We have observed the size effect at other temperatures.
It also appears when $\xi \sim L/10$.

The size effects enhance the correlation length.
The skewed periodic boundary conditions are imposed in this paper.
Boundary effects produce correlation of a spin with itself.
Therefore, the correlation is always overestimated by the size effects.
Then, the correlation length is also overestimated.
The spin-glass correlation length of $L=9$ reaches $\xi_\mathrm{sg}=6$
 in the equilibrium limit as shown in Fig. \ref{fig:xi2}(a).
In a lattice of $L=9$ the largest distance is $L/2=4.5$.
The correlation length exceeds this limit.
Therefore, such a value does not have a physical meaning at all.
This is a consequence of the size effect.
One may think that the size effect appears when the correlation length
reaches the system size.
However, it appears when $\xi$ reaches $L/10$.
Finite-size effects are very strong in spin glasses.
We must pay much attention to the size effects on numerical data.

These lines of evidence may answer the question why the nonequilibrium 
relaxation method can handle the spin-glass problems.
Relaxation processes in the scaling region know the critical phenomenon 
from the very early time steps.
There is a characteristic length that we can observe the critical behavior.
This is $\xi=0.5$ lattice spacings in the present simulations.
This length is surprisingly small.
This is because resolution of our simulations is very sharp.
It is an advantage of the nonequilibrium relaxation method.
Critical relaxation continues until the finite-size effects appear.
It is a time when $\xi$ reaches $L/10$.
If the temperature is off the transition temperature,
critical relaxation continues until the correlation length reaches 
its equilibrium finite value at the temperature.
What is slow in spin glasses is a relaxation process after the finite-size
effects appear.
For example, $\xi_\mathrm{sg}$ of $L=9$
in Fig. \ref{fig:xi2}(a) deviate from the size-independent line at $t\sim 30$.
It reaches the equilibrium limit at $t=10^5$.
Computations before this time are discarded in the equilibrium simulations.
On the other hand, those before the size effects appear ($t\sim 30$) are
utilized in the nonequilibrium relaxation method.
Therefore, the critical phenomenon of spin glasses is easily observed
if we take the nonequilibrium relaxation approach.
Another advantage of this approach is that the system size can be 
considered infinity.
The size effects are very strong and strange in spin glasses as shown in
this paper.
We can get rid of these effects by this method.
These are the reasons why the nonequilibrium relaxation method 
is successful in the spin-glass problems.

\section{Summary and Discussion}
\label{sec:discussion}

In this paper we made clear that the simultaneous spin-glass 
and chiral-glass transition occurs in the
$\pm J$ XY model in three dimensions.
We have eliminated ambiguity of the scaling analysis by introducing a new
criterion: correct results are independent of the temperature range of data
used in the scaling.
The results are checked and confirmed by the relaxation
functions of the spin-glass susceptibility, the Binder parameter 
and the spin-glass correlation length.
They are consistent with each other.
The evaluated transition temperature is higher than other 
estimations,\cite{Lee,Granato}
while the critical exponents are consistent.

The transition temperature of the spin-glass transition and 
that of the chiral-glass transition agree well within the error bars.
A ratio of the static critical exponents, $\gamma/\nu$, takes almost 
the same value between two.
They are $\gamma/\nu=2.0(2)$ for the spin and
$\gamma_{\kappa}/\nu_{\kappa}=1.9(2)$ for the chirality.
This agreement suggests that two transitions are qualitatively equivalent.
The dynamic exponent $z$ of the spin-glass obtained by the Binder parameter
and that obtained by the correlation length becomes consistent when
$T_\mathrm{sg}=0.47$.
This temperature is an upper bound of the error bar and is very close to
the mean value of the chiral-glass transition temperature.

The spin-glass susceptibility and the Binder parameter exhibit normal
behavior of the second-order phase transition.
Both exhibit algebraic divergence at the transition temperature.
An exponent of the divergence is strongest at the transition temperature.
Contrarily, those of the chiral-glass transition show strange behavior.
The chiral-glass susceptibility diverges algebraically at the transition 
temperature. 
However, 
an exponent of the susceptibility takes a maximum value at a lower temperature.
Relaxation of the Binder parameter shows no divergence.
It just remains zero in the infinite-size system.
If the transition is driven by the chirality degrees of freedom,
the Binder parameter should exhibit a kind of critical behavior
as the spin-glass one exhibits.
Relaxation functions of the chiral-glass correlation length 
exhibit the same behavior as the spin-glass one.
They begin the algebraic divergence when $\xi_\mathrm{sg/cg}\sim 0.5$ lattice
spacings and stop it at $\xi_\mathrm{sg/cg}\sim L/10$ 
when the finite-size effects appear.
The size effects appear later in the chiral-glass because
$\xi_\mathrm{cg}<\xi_\mathrm{sg}$.
Kawamura and Li\cite{KawamuraXY3} commented that the spin-chiral separation
occurs at $T=T_\mathrm{cg}$ at $t=10^5$ Monte Carlo steps in the 
standard Metropolis dynamics.
We have carried out simulations up to $t=10^6$ steps
and found no sign of the separation in behaviors of the infinite-size system.

The lines of evidence can be explained if one considers that the 
glass transition is driven by the spin degrees of freedom.
The chirality is defined by spin variables.
It trivially freezes if the spin freezes.
A critical property may be observed by some quantities of the chirality,
but it may not in other quantities.
The former example is the chiral-glass susceptibility and
the latter one is the Binder parameter.
This is because the chirality is the secondary degrees of freedom in 
this phase transition.
Every analysis and the results are consistent in the spin-glass transition.
This is a strong support for our argument 
that the transition is driven by the spin.

The present analyses are based upon numerical simulations.
The results always have error bars.
We cannot prove that the spin-glass transition temperature is exactly
same with the chiral-glass one.
Our estimations for the transition temperatures agree within error bars.
However, the mean value of the chiral-glass transition temperature
is a little larger.
In order to make a distinction between two transition temperatures
 we must carry out simulations 400000 times as large as the present ones.
The estimation is as follows.
If the mean values are correct, $\chi_\mathrm{sg}(t)$ should exhibit
converging behavior when $T=0.467$,
while $\chi_\mathrm{cg}(t)$ remains diverging.
We observed converging behavior of $\chi_\mathrm{sg}(t)$
at $t=10^4$ when $T=0.56$ as shown in Fig. \ref{fig:chi1}(a).
Therefore, the converging behavior is observed at 
$t=3.3\cdot 10^8$ when $T=0.467$ because $\tau\propto (T-0.455)^{-4.8}$.
The size effects appear when $\xi\sim L/10$ and 
$\xi_\mathrm{sg}(t)=0.4t^{1/5.6}$.
Then, a linear lattice size $L=130$ is necessary to obtain
size-independent data until $t=3.3\cdot 10^8$.
Therefore, $3.3\cdot 10^3 \times (130/39)^3\simeq 400000$ times larger
simulations are necessary.
It is almost impossible to make the distinction by raw relaxation functions.

Why the spin-glass transition has not been observed until quite recently?
The first reasonable answer we consider is strong finite-size effects.
As shown in Fig. \ref{fig:xi2} the size effects appear when the correlation
length reaches $L/10$.
Physically-meaningful spin-glass correlation functions are restricted within
this length scale.
Those outside this scale are plagued by the size effects.
Since the susceptibility collects all the correlation functions, 
it is also under the strong influence of the size effects.
Most of information extracted from the susceptibility may be unphysical one
if the lattice size is small.
Observations of the spin-glass susceptibility by
equilibrium simulations with small lattices
may suffer from this difficulty.
The second reason is the boundary conditions.
Most simulations employ the periodic boundary conditions.
They may produce a domain-wall spin state, or a spiral spin state in
continuous spin models.
Particularly, when the system possesses frustration,
nontrivial states may be produced.
On the other hand, the nonequilibrium relaxation method 
achieves the infinite-size system.
The obtained data are always independent of the finite-size effects and 
the boundary conditions.
True physical properties are easily extracted.

Recently, the spin-glass transitions are observed by a finite-size
scaling analysis on the spin-glass correlation length.\cite{Lee}
The quantity is considered to be less influenced by the size effects.
In the present paper we have also obtained relaxation functions of the
correlation length.
It is evaluated directly from the correlation functions.
Unphysical garbage data of correlations are not contained in our estimations.
Therefore, the spin-glass transition is clearly exhibited.
It is also found that the correlation length exhibits critical behavior 
from the first step.
Analyses on the correlation length will become a standard approach
in the spin glass problems.

\acknowledgments
The authors would like to thank Professor Fumitaka Matsubara
for guiding them to the spin-glass study.
One of the authors (TN) would like to thank Professor Hikaru Kawamura
and Dr. Hajime Yoshino for fruitful discussions, and
Professor Nobuyasu Ito and Professor Yasumasa Kanada 
for providing him with a fast random number generator RNDTIK.
He also would like to thank Professor Ian Campbell for pointing out
consistency of our data with those of Ref.\cite{KawamuraXY3}.
Simulations are partly carried out at the Supercomputer Center, ISSP, 
The University of Tokyo.
This work is supported by Grand-in-Aid for Scientific Research from
the Ministry of Education, Science, Sports and Culture of Japan (No. 15540358).

\end{document}